\documentclass[10pt]{article}

\def\a{\alpha}

\def\la{\lambda}

\def\ze{\zeta}

\def\th{\theta}

\def\De{\Delta}
\def\La{\Lambda}
\def\phi{\varphi}

\def\pd{\partial}
\def\d{{\rm d}}       

\def\({\left(}
\def\){\right)}
\def\[{\left[}
\def\]{\right]}
\def\~#1{\widetilde #1}
\def\.#1{\dot #1}
\def\^#1{\widehat #1}

\def\beq{\begin{equation}}
\def\eeq{\end{equation}}

\def \ov{\over}
\def \lb{\label}

\def\={\, =\, }

\def\dst{\displaystyle}

\def \sy {symmetry}
\def \sys {symmetries}
\def \so {solution}
\def \eq{equation}

\def\lss{$\la$-\sys}

\def \q{\quad}
\def \sk{\medskip}
\def \ni{\noindent}
\def\vf {vector field}
\def\R{{\bf R}}

\date{}

\begin{document}

\title{{\bf Reduction of systems of first-order differential equations via 
$\Lambda$-symmetries}}

\author{
   Giampaolo Cicogna\thanks{Email: cicogna@df.unipi.it} 
   \\~\\
Dipartimento di Fisica ``E.Fermi'' dell'Universit\`a di Pisa\\
  and  Istituto Nazionale di Fisica Nucleare, Sez. di Pisa \\~\\
Largo B. Pontecorvo 3, Ed. B-C, I-56127, Pisa, Italy  }

\maketitle

\ni{\bf Abstract}

The notion of $\la$-symmetries, originally introduced by C. Muriel and J.L. Romero, 
is extended to the case of systems of first-order ODE's (and of dynamical systems in particular).
It is shown that the existence of a symmetry of this type produces a reduction of 
the differential equations, restricting the presence of the variables
involved in the problem. The results are compared with the case of standard 
(i.e. exact) Lie-point symmetries and are also illustrated by some examples.

\bigskip \ni

\bigskip \ni

\section{Introduction}

It is a well known property that if an ordinary differential \eq\ (ODE) 
admits a Lie point-\sy , then the order of the \eq\ can be lowered  by one 
(see e.g. \cite{Ol}). The notion of $\la$-\sy\ has been introduced in 2001 by
Muriel and Romero \cite{MR1,MR2} with the main purpose of obtaining this
reduction even in the absence of standard Lie \sys . The idea consists in
introducing a suitable modification of the prolongation rules of the \vf\
in such a way that the lowering procedure  still works,
even if $\la$-\sys\ are not \sys\ in the proper sense, as they do not map
in general \so s into \so s.

$\la$-Symmetries are related to \sys\ of integral exponential type \cite{Ol,MR1,MR2}, 
to hidden and to some classes of potential \sys\ (see \cite{MR2,ASG,GMM} 
and references therein).
The meaning of $\la$-prolongation  has been clarified 
(together with  a possible generalization of the procedure) by means 
of classical theory of characteristics of \vf s \cite{PS}.
$\la$-Symmetries have been extended to partial differential \eq s \cite{GM,CGM} (and
called in that context $\mu$-\sys ), and also interpreted in terms of a 
deformed Lie derivative in a more geometrical approach \cite{PMo}. A nontrivial 
relationship with nonlocal \sys\ has been recently pointed out 
\cite{DCF}; an interpretation in terms of  appropriately defined changes of reference frames
has been also proposed \cite{Gfr}. 
For the implications of $\la$- and $\mu$-\sys\ in Noether-type conservation rules see 
\cite{MRO,CG}.

In the case of first-order ODE's, Lie \sys\ cannot lower the order of the 
\eq s, but they  provide a sort of  ``reduction'' of the complexity of the 
system, or -- more precisely -- a reduction of the number of the involved 
variables (see \cite{Ol},  Theorem 2.66). In this paper, we will restrict precisely to the 
case of {\it systems} of {\it first-order} ODE's (with usual regularity 
and nondegeneracy assumptions: see e.g. \cite{Ol,BA}), and of dynamical 
systems (DS) in particular, where the
application of \lss\ requires some attention and where they exhibit some
relevant peculiarities. We shall prove that some forms of reduction are
allowed also in these cases.

An application of \lss\ to systems of ODE's has been already considered  in a
particular case \cite{MR3}; in the present paper we want to examine more general
situations.

\section{Systems of ODE's}

Let us recall first of all that in the case of a {\it single} dependent variable $u=u(t)$
(we shall always denote by $t\in\R$ the independent variable, according to
its natural interpretation as the time variable in the case of DS), the 
first-order $\la$-prolongation $X_\la^{(1)}$ of a \vf\ $X$
\beq\lb{X0} X\=\tau(t,u){\pd\ov{\pd t}}+\phi(t,u){\pd\ov{\pd u}}\eeq
is defined as
\beq\lb{defXl} X_\la^{(1)}\=X^{(1)}+\la\, Q{\pd\ov{\pd\.u}}\eeq
where $X^{(1)}$ is the {\it standard} prolongation \cite{Ol,BA}, $\.u=\d u/\d t$,  
$\la\=\la(t,u,\.u)$ is an arbitrary $C^\infty$ function, and
$Q\=\phi-\tau \,\.u$. 

Considering systems of \eq s, and then $q>1$ dependent variables 
$u_a=u_a(t)$, the natural extension of definition (\ref{defXl}) is
\beq\lb{defXL} X_\La^{(1)}\=X^{(1)}+(\La Q)_a{\pd\ov{\pd\.u_a}}\eeq
where the sum over $a=1,\ldots,q$ is understood, with
\beq\lb{X} X\=\tau{\pd\ov{\pd t}}+\phi_a{\pd\ov{\pd u_a}}\, , \q\q\q 
Q_a\=\phi_a-\tau\,\.u_a\eeq
and where now $\La$ is a $q\times q$ matrix of $C^\infty$ functions depending on 
$t,u_a,\.u_a$. The case $\La=\la\,I$ is the one considered, in the context 
of DS and also for systems of ODE's of any order, by Muriel and Romero \cite{MR3}.

Given a system of $q$ first-order ODE's (we shall assume for simplicity 
that the number of the \eq s is the same as the number of dependent 
variables $u_a(t)$)
\beq\lb{Fa} F_a(t,u_b,\.u_b)\=0 \q\q\q a,b=1,\ldots,q\eeq
we shall say that this system is $\La$-{\it symmetric} under a \vf\ $X$ if there 
is a matrix $\La$ such that
\beq\lb{Lasy} X^{(1)}_\La\,F_a\big|_{F_a=0}\=0 \ .\eeq
It is clear from (\ref{defXL}) that $\La$ is not uniquely defined: indeed, 
for any matrix $R$ such that $RQ=Q$ then also $\La'=\La R$ 
satisfies the above condition. This 
arbitrariness in the definition of $\La$, far from being disturbing, may 
be useful in practice, as it allows the choice of the more convenient 
matrix $\La$ in view of the given problem.

We shall say that the system (\ref{Fa}) is $\La$-{\it invariant} under
$X$  if there is a matrix $\La$ such that
\beq X^{(1)}_\La\,F_a\=0 \ .\eeq

It is not too restrictive to assume that the system of ODE's we are going to consider can 
be put into a $\La$-invariant form. Indeed, extending to $\La$-\sys\ a well known result
\cite{Ol,PS}, it can be shown that if a system is $\La$-symmetric, then there exists
a $q\times q$ matrix $A=A(t,u,\.u)$ such that 
$$X^{(1)}_\La\,F_a\=A_{ab}F_b \ .$$
It is now enough to prove, applying standard arguments (cf.  e.g. \cite{CLlib,CDW,CL}), 
the existence of  some some $q\times q$ invertible matrix $S$, 
possibly depending on $t,u,\.u$, such that 
$$X^{(1)}_\La\,S+S\,A\=0  $$
and the (locally) equivalent system $G_a\equiv S_{ab}F_b\=0$ turns out to be $\La$-invariant.
We will consider in the following, unless otherwise stated,
only $\La$-invariant systems.

The matrix $\La$ plays the role of an additional  ``unknown'' in the determining \eq s
which are deduced from the $\La$-invariance condition
\beq\tau{\pd F_a\ov{\pd t}}+\phi_b{\pd F_a\ov{\pd u_b}}+\(\phi_b^{(1)}+(\La Q)_b\){\pd F_a\ov{\pd \.u_b}}\=0\eeq
where $\phi_b^{(1)}$ is the coefficient of the standard first-order prolongation, and which clearly strongly depend  on the explicit form of the functions $F_a$. For instance, in the case where $F_a=\.u_a-f_a(t,u)$, i.e. the case of dynamical systems (see Section 3), these \eq s are
$$\{\phi\, ,\, f\}_a\={\pd\ov{\pd t}}(\phi_a-\tau\,f_a)-{\pd\tau\ov{\pd u_b}}f_af_b+\La_{ab}Q_b \ , $$
where $\dst{\{\phi\, ,\, f\}_a=\phi_b{\pd f_a\ov{\pd u_b}}-f_b{\pd \phi_a\ov{\pd u_b}}}$, which clearly become 
\beq\lb{LPTI} \{\phi\, ,\, f\}_a\=(\La\,\phi)_a \eeq
in the case of autonomous systems and time-independent \vf s $X$ with $\tau\equiv 0$
(see \cite{cglib}).

\sk
Let us now introduce  ``\sy -adapted'' coordinates $w_a$
(sometimes also called  canonical coordinates) characterized by the property of
being invariant under the  action of the \vf\ $X$:
\beq\lb{defwa} X\,w_a\=\tau{\pd w_a\ov{\pd t}}
 + \phi_b{\pd w_a\ov{\pd u_b}}\=0 \ ;\eeq
they are obtained through the associated characteristic  \eq  s 
$${\d t\ov{\tau}}\={\d u_a\ov{\phi_a}} \ .$$
In this way we introduce  exactly $q$ new variables $w_a$; one at least 
of these, say $w_q$, will depend explicitly on $t$, and we will choose 
this as the new independent variable and call it $\eta$. In particular, 
if $\tau\equiv 0$, we can choose $\eta=t$. As $(q+1)$-th 
variable, which will be called $z$, we will take the  coordinate ``along 
the action of $X$'', i.e. such that $X\,z=1$. Summarizing, the new set of 
variables is 
\beq\lb{wz} \eta,\,w_\a(\eta),\, z(\eta),\ \q\q\q 
\a=1,\ldots,q-1  \eeq
(among these, $w_\a$ and  $w_q\equiv\eta$ are invariant under $X$) 
and clearly do {\it not} depend on $\La$.

We have now to write the given \vf\ $X$ and its first $\La$-prolongation (\ref{defXL})
in terms of these coordinates. We get first $X\=X^{(1)}\=\pd /\pd z$, and we then find
that eq. (\ref{defXL})  takes the form 
\beq\lb{XtLa} X_\La^{(1)}\= {\pd\ov{\pd z}}+M_\a{\pd\ov{\pd 
w'_\a}}+M_q{\pd\ov{\pd z'}}
\eeq
where   $w_\a'=\d w_\a/\d\eta,\, z'=\d z/\d\eta$ and
(here and in the following the sum will be always understood over the repeated indices 
$\a=1,\ldots,q-1$ and $a=1,\ldots,q$; $D_t$ is the total derivative)
\beq\lb{ma} M_\a\=(D_t\eta)^{-2}\Big(D_t\eta{\pd w_\a\ov{\pd 
u_a}}-D_tw_\a{\pd\eta\ov{\pd u_a}}\Big)(\La Q)_a \eeq
\beq\lb{mq} M_q\=(D_t\eta)^{-2}\Big(D_t\eta{\pd z\ov{\pd
u_a}}-D_tz{\pd\eta\ov{\pd  u_a}}\Big)\ (\La Q)_a \ .\eeq
In particular, if $\tau\equiv 0$, and then with 
$\eta=t$, we have more simply
$${\pd w'_\a\ov{\pd\.u_a}}\={\pd w_\a\ov{\pd u_a}} \q\q , \q\q
{\pd z'\ov{\pd \.u_a}}\={\pd z\ov{\pd u_a}}$$
and
\beq\lb{mt0}M_\a\={\pd w_\a\ov{\pd u_a}}(\La\,Q)_a\q\q , \q\q M_q\={\pd 
z\ov{\pd u_a}}(\La\,Q)_a \ . \eeq
The above expressions (\ref{ma},\ref{mq}) can be obtained either by direct calculation expressing 
by the chain rule the operators 
$\pd/\pd\.u_a$ in terms of $\pd/\pd w'_\a,\,\pd/\pd z'$, or -- more elegantly -- starting from
the algebraic relation
\beq\lb{XD} [X^{(1)}_\La,\,D_t]\=-D_t(\tau)D_t+(\La Q)_a{\pd\ov{\pd
u_a}}\eeq
which can be easily proved and generalizes to $\La$-\sys\ other similar known identities \cite{MR1,MR2}. From this, one directly gets indeed
$$X_\La^{(1)}(w'_\a)\=X_\La^{(1)}\Big({D_tw_\a\ov{D_t\eta}}\Big)\=
{X_\La^{(1)}(D_tw_\a)(D_t\eta)-(D_tw_\a)X_\La^{(1)}(D_t\eta)\ov{(D_t\eta)^2}}\=$$
\beq\lb{reob}\hskip-.5cm=(D_t\eta)^{-2}\Big(D_t\eta{\pd w_\a\ov{\pd u_a}}-
D_tw_\a{\pd\eta\ov{\pd u_a}}\Big)(\La Q)_a\=M_\a\eeq
thanks to $X^{(1)}_\La w_\a=X w_\a=0$, $X^{(1)}_\La\eta=X \eta=0$;
similarly for $X_\La^{(1)}(z')$.

\sk
It can be interesting to point out that eq. (\ref{reob}) 
puts in clear evidence the difference with respect to exact \sys: indeed,
starting from the $q$ ($X$-invariant) variables $w_\a,\, \eta$ one obtains
$q-1$ first-order differential quantities $w'_\a$ which are
invariant under $X^{(1)}$, but in general {\it not} under $X_\La^{(1)}$. 

\sk
In turn, the given system of differential \eq s will take the form
(we will use the $\,\~\cdot\, $ to denote the expressions 
in the new variables)
\beq\lb{Ft}\~F_a(\eta,w_\a,w'_\a,z,z')\=0\eeq
and the condition of its $\La$-invariance under $X$ now becomes 
\beq\lb{Xtinv}{\pd\~F_a\ov{\pd z}}+M_\a{\pd \~F_a\ov{\pd
w'_\a}}+M_q{\pd\~F_a\ov{\pd  z'}}\=0 \ .\eeq
This allows us to state the following first form of reduction:

\sk
\ni {\bf Theorem 1}. {\it If the system (\ref{Fa}) is $\La$-invariant under a 
\vf\ $X$, then, once written in the \sy-adapted coordinates 
$\eta,w_\a,w'_\a,z,z'$, it 
turns out to depend on only $2q$ quantities (instead of  $2q+1$): i.e. 
on the $q$ variables $w_\a,\,\eta$ and on   other $q$ first-order differential $\La$-invariant 
quantities $\ze_a=\ze_a(\eta,z,w_\a,w'_\a,z')$ which are obtained from the  characteristic  \eq s
\beq\lb{fodi} \d z\={\d w'_\a\ov{M_\a}}\={\d z'\ov{M_q}}\eeq
coming from condition (\ref{Xtinv}).}

\sk
Examples 1 and 2 will illustrate this result.

\section{The case of Dynamical Systems}

Let us now consider the particularly important case of the dynamical systems, 
i.e. the systems of first-order ODE's which are written ``in explicit form'':
\beq\lb{DS} \.u_a\=f_a(t,u) \ .\eeq

Clearly, once  \sy-adapted coordinates are introduced, the system becomes ``automatically'' 
a function of the $2q$ quantities $w_\a,\eta,\zeta_a$, as granted by Theorem 1. 

But it can be preferable or more convenient (e.g. in view of the physical interpretation in terms of  ``evolution'' problem, or also if the explicit expression of the $\zeta_a$ is not known\footnote{If  one is interested to know ``a 
priori'' the expressions of the $q$  differential $\La$-invariant 
quantities $\ze_a=\ze_a(\eta,z,w_\a,w'_\a,z')$, one has to 
express $M_\a,M_q$ in terms of $w_\a,\eta$ in order to solve (\ref{fodi}). }) to adopt a different point of view, i.e. to preserve the  form of the system as an explicit DS, i.e. to rewrite it as follows
\beq\lb{DStd} w'_\a\=\~f_\a(\eta,w,z) \eeq
\beq\lb{DStz} \,\,\,z'\,\,\=\~f_q(\eta,w,z)\ . \eeq
and to look for the dependence on $z$ of the r.h.s. This point of view will be elucidated by Examples 3 and 4. 

Recalling the expression (\ref{XtLa}) of the first $\La$-prolongation of $X$, we then easily deduce in this case:

\sk
\ni{\bf Theorem 2.} {\it If a DS is $\La$-invariant under $X$,  the dependence on $z$ of the r.h.s. of eq.s (\ref{DStd},\ref{DStz}) is given by 
$${\pd \~f_\a\ov{\pd z}}\=M_\a\q\q ;\q\q {\pd \~f_q\ov{\pd z}}\=M_q \ .$$
Then, if for some $\overline{\a}$ one has $M_{\overline{\a}}=0$,  the corresponding 
$\~f_{\overline{\a}}$ does not depend on $z$. If $M_\a=0$ for all $\a=1,\ldots,q-1$,  
then only $\~f_q$ depends on $z$ and the system splits into a system for the $q-1$ 
variables $w_\a=w_\a(\eta)$ and the last \eq\ (\ref{DStz}) which is an ODE for 
the variables $z$ and $\eta$.}

\sk
It is useful to compare the situation covered by Theorems 1 and 
2 with the case of exact \sy : the difference is that in the  case of exact \sy\
all terms of eq.\,(\ref{Ft}) are independent of $z$; the same is true for all the terms at the r.h.s. 
of  (\ref{DStd},\ref{DStz}):
then, in this case,  the last \eq\ for $z$ and $\eta$ turns out to be a 
quadrature, as is well known \cite{Ol}.

Clearly, if $\La=0$ i.e. if  $X$ is an exact \sy , then $M_\a=M_q=0$. 
Conversely, it can be shown that if $M_\a=M_q=0$ then the \sy\ $X$ is 
exact. This is particularly clear in the case $\tau\equiv 0$ (and then 
$\eta=t$): indeed, in this case the conditions $M_\a=M_q=0$ can be 
written (see (\ref{mt0})) ${\bf J}_{ab}(\La\,Q)_b=0$ where ${\bf J}$ is the (invertible !) 
Jacobian matrix of the transformation from $u_a$ to $w_\a,z$. Then 
$\La\,Q=0$, which is the same as $\La=0$ (recall that $\La$ is not 
uniquely defined).

Notice in particular that the term $(\La\,Q)_a$ 
appearing in the expressions (\ref{ma},\ref{mq}), when written in the new 
coordinates, becomes
$$(\~\La\,\~Q)_a\=\~\La_{aq}$$
indeed $\~Q\equiv(0,0,\ldots,1)$. 
This shows that only the last column of $\~\La$ is relevant.

\sk
Finally, let us recall  the following result:

\sk
\ni{\bf Theorem 3.} (Muriel-Romero \cite{MR3}) {\it If $\La=\la\,I$, then 
$M_\a=0$ for all $\a=1,\ldots,q-1$, and the conclusion of the last part of 
Theorem 2 holds.}

\sk
Indeed, from $X\,w_\a=X\,\eta=0$, $X\,z=1$ and the definition of $Q$, one easily deduces
$$Q_a{\pd w_\a\ov{\pd u_a}}\=-\tau\,D_tw_\a \q\q ; \q\q Q_a{\pd \eta\ov{\pd u_a}}\=-\tau\,D_t\eta$$
and
$$Q_a{\pd z\ov{\pd u_a}}\=-\tau\,D_t z+1$$
hence, in the case $(\La\,Q)_a=\la Q_a$ considered in \cite{MR3}, one gets
$$M_\a\=0 \q\q ;\q\q M_q\=(D_t\eta)^{-2}\la \ .$$
Notice also that (\ref{XD}) becomes in this case 
$$  [X^{(1)}_\la,\,D_t]\=-D_t(\tau)D_t+ \la Q_a{\pd\ov{\pd
u_a}}\=-D_t(\tau)D_t+\la X-\la\tau D_t \ . $$

\section{Examples}

\sk\ni
{\bf Example 1}. This is a very simple example, which can provide a clear illustration of Theorem 1. Consider  any system $F_a(t,u_1,u_2,\.u_1,\.u_2)=0\ (a=1,2)$ of two first-order ODE's for the variables $u_1=u_1(t)\,,u_2=u_2(t)$ and consider the \vf\ 
$$X\={\pd\ov{\pd u_2}}\ .$$
It is easily seen that if one chooses
$$ \La\=\pmatrix{0 & 1\cr 0& 1}$$
then, with our notation, $w_1=u_1,\, \eta=t$ and $z=u_2$; eq. (\ref{ma},\ref{mq}) give $M_1=M_2=1$ and therefore from (\ref{fodi}) $\zeta_1=\.w_1-z=\.u_1-u_2,\, \zeta_2=\.z-z=\.u_2-u_2$. Then,  $\La$-invariance under $X$ gives that the $F_a$  depend only on the quantities $t,\,\.u_1-u_2,\,\.u_2-u_2$, in agreement with Theorem~1. Extension to more than 2 variables $u_a$ is immediate.

\sk\ni
{\bf Example 2}. Consider a
system of ODE's for the two variables $u_1=u_1(t),\, u_2=u_2(t)$ of the form
$$h(s_1,s_2)\,(\.u_1-u_1u_2)+a(t)(u_1^2+u_2^2)u_1+b^2(t)(u_1^2+u_2^2)u_2\=0$$
$$\hskip -.3cm h(s_1,s_2)\,(\.u_2+u_1^2)+a(t)(u_1^2+u_2^2)u_2-b^2(t)(u_1^2+u_2^2)u_1\=0$$
where  $h$ is a function of $s_1=u_1\.u_1+u_2\.u_2,\, s_2=u_1\.u_2-\.u_1u_2+u_1^3+u_1u_2^2$ 
and where $a(t),\,b(t)$ are arbitrary functions of $t$; it is clearly not symmetric under the 
rotation operator
$$X\=u_2{\pd\ov{\pd u_1}}-u_1{\pd\ov{\pd u_2}}$$
(unless $h\equiv 0$), however it turns out to be $\La$-symmetric (but not $\La$-invariant) 
under rotations if $\La=\la\,I$ with
$\la=u_2$. Indeed, e.g., one has $X^{(1)}_\La\,h=~0$, $X^{(1)}_\La\,(\.u_1-u_1u_2)=(\.u_2+u_1^2)$, etc. Introducing  \sy-adapted coordinates, which are $w_1=r=(u_1^2+u_2^2)^{1/2},\, \eta=t,\, z=\th$, with obvious notations, the system becomes
$$\hskip-1.7cm  \.r\, h(r\.r,r^2(\.\th+r\cos\th))+a(t)r^3\=0$$
$$(\.\th+r\cos\th)\,h(r\.r,r^2(\.\th+r\cos\th))-b^2(t)r^2\=0$$
which turns to be $\La$-invariant under $X=\pd/\pd\th$ with $\la=r\sin\th$. As expected, thanks to Theorem 1, this system contains only the four quantities $r,t$ and $\ze_1=\.r,\ze_2=\.\th+r\cos\th$. 

If, e.g., $h=s_2$, the system can be also put in the explicit form of a DS:
$$\.r=\pm(a(t)/b(t))r \q\q\ \.\th\=\pm b(t)-r\cos\th$$
and -- as a consequence -- according to Theorems 2 and 3, one (and only one) of the above \eq s does not contain $z$ (here: $\th$). Then the system can be easily solved.

\sk\ni
{\bf Example 3}. Consider any DS for $u_a=u_a(t), \, a=1,2,3$, of the form
$$\.u_1\=h_1(t,w_1,w_2)+(a-3b)u_2u_3+b\,u_3^3+h_2(t,w_1,w_2)\,u_3+h_3(t,w_1,w_2)\,u_3^2$$
$$\hskip-4.9cm  \.u_2\=h_2(t,w_1,w_2)+2u_3h_3 (t,w_1,w_2)+au_3^2$$
$$\hskip-7.6cm  \.u_3\=cu_3+2h_3(t,w_1,w_2)$$
where $a,b,c$ are constants and $h_a$ are functions of $t,\, w_1=2u_2-u_3^2,$
$w_2=3u_1-3u_2u_3+u_3^3$. Systems of this form are
$\La$-invariant under the \vf
$$X\=u_2{\pd\ov{\pd u_1}}+u_3{\pd\ov{\pd u_2}}+{\pd\ov{\pd u_3}}\q\q {\rm with} \q\q \La=
\pmatrix{0 & 0 & (a-3b)u_2\cr 0&0&(2a-c)u_3\cr 0&0&c} \ .$$
The $X$-invariant quantities are just $w_1,w_2$, together with $\eta=t$. The coefficients $M_\a,M_q$ (see eq.s (\ref{ma},\ref{mq})), with $z=u_3$, are
$$M_1\=4(a-c)z\q\q, \q\q M_3=c\q\q,$$
$$M_2\=3(a-3b-c)u_2-6(a-c)u_3^2=3(a-3b-c)(w_1+z^2)/2-6(a-c)z^2\ . $$
The characteristic  \eq s (\ref{fodi}) can then be easily solved to obtain the three first-order differential $\La$-invariant quantities $\ze_1=\.w_1-2(a-c)z^2,\,\zeta_2=\.w_2-(3/2)(a-3b-c)w_1z-(3/2)(a+b-c)z^3,\,\zeta_3=\.z-c\,z$. Direct calculation shows  that this system becomes
$$\ze_a-g_a(t,w_1,w_2)\=0$$
where $g_1=2h_1,\,g_2=3h_1-3h_3w_1,\,g_3=2h_3$, and then contains only the quantities $t,w_1,w_2,\ze_a$, in agreement with Theorem 1. If instead one prefers to write the system as an explicit DS, then it is
$$\hskip-4.1cm \.w_1\=2(a-c)z^2+g_1(t,w_1,w_2)$$
$$\.w_2\={3\ov 2}(a-3b-c)w_1z+{3\ov 2}(-a-b+c)z^3+g_2(t,w_1,w_2)$$
$$\hskip-4.8cm\.z\,\=c z+g_3(t,w_1,w_2)\ .$$
Now, if $a=c$, the first \eq\ does not contain $z$; if $a=c$ and $b=0$, only the third \eq\  contains $z$, in agreement with Theorem 2. If $a=c=0$, then only the second \eq\  contains $z$ and ``plays the role'' of the $q-$th \eq\ in our notation. The case $a=b=c=0$ is of course the case of exact \sy\ $\La=0$.

\sk\ni
{\bf Example 4}. This is an example with non-autonomous DS and  \vf\ $X$ with $\tau\not=0$ and therefore $\eta\not=t$.
Consider the DS for $u_a=u_a(t), \, a=1,2,3$,
$$\hskip-4.5cm\.u_1=t+h_1(s,w_1,w_2)\,\exp(-\la_1 t)$$
$$\hskip-4.5cm\.u_2=1+h_2(s,w_1,w_2)\,\exp(-\la_2 t)$$
$$\.u_3=u_2+h_2\exp(-\la_2 t){1-\exp(u_2\big(\la_2-\la_3)\big)\ov{\la_3-\la_2}}+h_3\exp(-\la_3 u_2)$$
where $h_a$ are nonvanishing functions of $s=u_2-t,\, w_1=u_1-t^2/2,\, w_2=u_3-u_2^2/2$. This
system is $\La$-invariant under
$$X\={\pd\ov{\pd t}}+t{\pd\ov{\pd u_1}}+{\pd\ov{\pd u_2}}+u_2{\pd\ov{\pd u_3}}$$ 
with $\La={\rm diagonal} (\la_1,\la_2,\la_3)$. We can choose as invariants under $X$  just $w_1,w_2$ and $\eta=u_2-t$, with $z=u_2$. It is now more useful to rewrite the system in these coordinates preserving its form of explicit DS, we get then
$$\hskip-4.5cm w'_1\=(h_1/h_2)\exp\((\la_2-\la_1)(z-\eta)\)$$
$$w'_2\=\Big({1-\exp(\la_2-\la_3)z\ov{\la_3-\la_2}}-z\Big)+(h_3/h_2)\exp\big(z(\la_2-\la_3)-\la_2\eta)\big)$$
$$\hskip-5.4cm z'\=(1/h_2)\exp\big(\la_2(z-\eta)\big) \ .$$
We see that if $\la_1=\la_2$, or $\la_2=\la_3$ (notice that the case $\la_2=\la_3$ is well defined), one of the above \eq s does not contain $z$ in agreement with Theorem 2; whereas if $\la_1=\la_2=\la_3$ only one \eq\ contains $z$, as stated by Theorem 3; the same happens also if $\la_1=\la_2=0$, according 
to Theorem 2. 

\section{Concluding remarks}

An interesting property which relates invariance with \lss\ is the following.
Consider the case of an autonomous DS $\.u_a=f_a(u)$ which is $\La$-invariant under a \vf\ $X$ of the form $X=\phi_a(u)\pd/\pd u_a$. If   $w=w(u)$  is any invariant under $X$, i.e. $\phi_a\pd w/\pd u_a$, 
then   its {\it Lie derivative  along} $f_a$, i.e.
$$D_t^{(f)}\,w\equiv f_a{\pd w\ov{\pd u_a}}$$
satisfies
$$X\,\(D_t^{(f)}\,w\)\=\Big(\phi_a{\pd\ov{\pd u_a}}\Big)\Big(f_b{\pd\ov{\pd u_b}}\Big)\,w\=\La_{ab}\phi_b{\pd w\ov{\pd u_a}} $$
hence $X\(D_t^{(f)}\,w\)=0$ if $\La=\la\,I$, having used the commutation rule (\ref{LPTI}) and the invariance property of $w$. On the other hand, eq. (\ref {reob}) gives directly, for $X$ of the above form,
$$X^{(1)}_\La\,(D_t\,w)\=(\La\phi)_a{\pd w\ov{\pd u_a}} \ .$$
The strong difference is that the latter result is purely algebraic, being a consequence of the relation (\ref{XD}), and expresses a property of the \vf\ $X$ which holds {\it independently} of the presence of any DS (i.e., of any choice of the functions $f_a$). The former result, instead, states  that the time evolution under the dynamics described by the DS $\.u_a=f_a$ of a quantity $w(u)$ which is invariant under a \vf\ $X$ {\it preserves} this invariance {\it even if $X$ is   not} a (standard) \sy\ of the DS; it is enough to require that $X$ is a $\la$-\sy\ of the DS. 

It can be noticed that the present statement, concerning Lie derivatives, can be suitably extended to the case of several \vf s $X$ for the given DS (see \cite{HW}, Prop. 2.1).

\sk
Several other aspects of \lss\ (and of all their generalizations as well) could be further investigated. 
Apart from their geometrical interpretation (see the papers quoted in the Introduction), their action on changes of coordinates should be better understood, as well as their general role in finding solutions of differential \eq s which do not admit standard \sys : see e.g. \cite{MR1,MR2,GMM,PS} and the references therein;   compare also, for instance, with \cite{IF}, for what concerns the problem of finding integrating factors for ODE's and its relationship with \sy\ properties.

\sk
It can be observed, finally, that any ODE $\De(t,u,\.u,\ddot u,\ldots)=0$ of 
arbitrary order~$>1$ can be transformed into a system of first-order 
ODE's, and therefore our results could be applied also to this case. This 
is true in principle: the only nearly obvious remark is that one has to
consider no longer \vf s of the form $X=\tau(\pd/\pd t)+\phi(\pd/\pd u)$ 
involving only the two variables $t$ and $u$, but also extended \vf s 
$X=\tau(\pd/\pd t)+\phi_{(0)}(\pd/\pd u)+\phi_{(1)}(\pd/\pd \.u)+
\phi_{(2)}(\pd/\pd \ddot u)+\ldots$. It is ``conceptually'' different to look for \vf s 
of the former or of the latter form; on the other hand, the ``concrete effect'' of 
the existence of a \sy\ is different in the two contexts (i.e., lowering the order 
in the case of the ODE's, and respectively reducing the presence of the involved variables in 
the case of first-order systems, as shown). This holds in particular for 
$\la$- and $\La$-\sys , where also the prolongation rules of 
the \vf s are markedly different in the two cases. To emphasize this different 
role of $\La$-\sys\ in the context of first-order systems, it should be perhaps more 
appropriate to call them $\rho$-\sys\ (where $\rho$ stands for ``reducing'', 
in contrast with $\la$, which could stand for ``lowering'').

\section*{Acknowledgments} It is a pleasure to thank Giuseppe Gaeta and Diego Catalano Ferraioli for detailed discussions and useful comments. Thanks are also due to one of the referees for his/her accurate and appropriate comments and suggestions, which helped me to improve the presentation of these results.

\sk\sk



\end{document}